\documentclass[final,5p,times,twocolumn,numbers]{elsarticle}
 
 \setcitestyle{square} 

\usepackage{xurl}

\usepackage{hyperref}
\usepackage{amssymb}
\usepackage{lipsum}

\usepackage{amsmath}
\usepackage{xcolor}
\usepackage{soul}
\journal{Physics Letters B}

\begin{document}
\begin{frontmatter}

\title{Moat hyperons in cold neutron stars}

\author[first,second]{Hristijan Kochankovski}
\author[first,second]{Angels Ramos}
\author[third,fourth]{Laura Tolos}
\affiliation[first]{organization={Departament de Fisica Quantica i Astrofisica (FQA), Universitat de Barcelona},
            addressline={Marti i Franques 1}, 
            city={Barcelona},
            postcode={08028},
            country = {Spain}}
            
 \affiliation[second]{organization={Institut de Ciencies del Cosmos, Universitat de Barcelona},
            addressline={Marti i Franques 1}, 
            city={Barcelona},
            postcode={08028}, 
            country={Spain}}   
            
\affiliation[third]{organization={Institute of Space Sciences (ICE, CSIC), Campus UAB}, addressline={Carrer de Can Magrans}, postcode={08193}, city ={Barcelona}, country = {Spain}}

\affiliation[fourth]{
organization={Institut d'Estudis Espacials de Catalunya (IEEC), Edifici RDIT, Campus UPC}, postcode={08860}, city={Castelldefels (Barcelona)}, country = {Spain}}

\begin{abstract}
We investigate the hyperonic equation of state within the non-linear derivative model that incorporates a momentum dependence on the interactions, with a special emphasis on properly establishing the conditions for hyperon appearance in neutron star matter. We demonstrate that hyperons can appear at finite momentum, forming a so-called ``moat" region, even when they are absent at lower momenta. Our study shows that this phenomenon significantly alters the composition and equation of state of hyperonic matter as compared to the cases when it is disregarded, highlighting the importance of accurately treating the momentum dependence of the baryon fields in dense matter. 

\end{abstract}


\end{frontmatter}

\section{Introduction}
\label{introduction}

The composition of matter at extreme densities, such as those found in the cores of neutron stars (NSs), has remained a fundamental question in nuclear astrophysics for over half a century. The inherent challenges in understanding the strong interaction at the low-energy scales relevant to these environments, tied to the non-perturbative nature of quantum chromodynamics (QCD), render the precise composition of neutron star matter model-dependent. While a majority of the models typically assume a composition of primarily nucleonic matter and leptons (electrons and muons), the densities within neutron stars can significantly exceed the nuclear saturation one, $\rho_0$\footnote{approximately a particle density of $0.16~\mathrm{fm^{-3}}$ or a mass density of around $2.7\cdot10^{14}~\mathrm{g/cm^{3}}$.}, by several  orders. This extreme compression leads to a substantial increase in the chemical potential of nucleons, potentially triggering the emergence of novel phases of matter, like the appearance of heavier baryons, such as hyperons, or even de-confined quark matter \cite{SchaffnerBielichCompactStarPhysicsbook2020}.

The notion that hyperons, baryons containing at least one strange quark, could be potential constituents of dense matter in neutron stars was first proposed in the seminal work of Ambartsumyan and Saakyan in 1960 \cite{AmbartsumyanHyperons1960}, and later investigated by many other authors (see the reviews of Refs.~ \cite{SedrakianPhysicsDenseHadronicMatter2007, OertelHyperonsNSSupernovas2016, ChatterjeeHyperonsExistNS2016, TolosStrangenessNucleiNS2020, LogotetaHyperonsNS2021, BurgioNSNuclearEoS2021}). The equations of state (EoSs) describing dense hyperonic matter can be derived using both microscopic and phenomenological theoretical frameworks. The inclusion of hyperons typically results in a softening of the EoS. This softening arises because the nucleon chemical potential at high densities reaches values that permit the conversion of energetic, degenerate nucleons into slow, less degenerate hyperons, thus relieving the Fermi pressure exerted by the nucleons in the EoS, and hence to a substantial decrease in the speed of sound. The softening of the EoS induced by hyperons stands in contrast with observations of the most massive pulsars \cite{DemorestShapiroMassiveNS2010,AntoniadisMassivePulsar2013,FonsecaNANOGRAV2016,CromartieRelativisticPulsar2020,RomaniFastestHeaviestNS2022}, which require the EoS of dense matter to be stiff enough to support their large gravitational masses. This discrepancy is commonly referred to as the ``hyperon puzzle" and various theoretical proposals have been made for its solution. On the one hand, within microscopic frameworks, a key approach involves the introduction of three-body forces between baryons (NNN, NNY, NYY, and YYY), which can provide additional repulsive contributions at high densities, thereby stiffening the EoS. On the other hand, phenomenological relativistic mean-field (RMF) models often introduce additional massive hidden strangeness vector mesons, which contribute with significant repulsive interactions when hyperons become abundant, again leading to a stiffer EoS. However, these phenomenological models often employ a relatively simple structure, particularly in their treatment of interactions that are generally momentum-independent.

In an attempt to address the limitations of momentum-independent interactions, a modification of the standard relativistic model, known as the Non-Linear Derivative (NLD) model, has been proposed \cite{ZimanyiEoSDerivativeScalarCoupling1990, TypelRMFGeneralizedDerivativeCouplings2003, TypelRMFMomentumDependentSelfEnergies2005}. 
This model explicitly
introduces couplings of mesons to derivatives of the baryon fields in the Lagrangian density, effectively leading to momentum dependent 
scalar and vector self-energies.
 The NLD model retains the computational simplicity characteristic of other RMF methods, while also exhibiting consistency with Dirac phenomenology, and can successfully describe nucleon-nucleus scattering using a momentum-dependent relativistic optical potential \cite{GaitanosNLDInteractionsRHD2009, GaitanosAntinucleonOpticalPotential2011,GaitanosEnergyDependentIsospinDynamics2012,GaitanosMomentumDependentMFNuclearmatterNS2013,GaitanosRMFDescriptionNNucleus2015}. The NLD model has also demonstrated success in describing the properties of nucleonic matter and atomic nuclei, and has recently been extended to include the appearance of hyperons in dense matter \cite{GaitanosMomentumDependentMFAntiY2021,ChorozidouMomentumDependenceInMediumPotential2024,ChorozidouNovelSolutionYPuzzle2024}. 
 
 This paper aims at reviewing and further investigating the hyperonic EoS within the NLD model, with a special focus on the conditions governing the appearance of hyperons in neutron star matter. We will modify the generally employed conditions to accommodate the possibility of hyperon production occurring at finite momentum, even in scenarios where  hyperons are not present at zero momentum. Our investigation into the momentum dependence of hyperon production could offer new insights into the resolution of the hyperon puzzle and the properties of ultra-dense matter.   

\section{Non Linear Derivative model}
As in the RMF models \cite{DuerrRelativisticEffectsNucForce1956,WaleckaTheoryCondensedMatter1974} and in any other RMF method, the NLD model assumes that the interaction between the baryons $\Psi = (N,\Lambda,\Sigma,\Xi)$ is accomplished by exchanging virtual mesons. The Lagrangian is written as:
\begin{equation}
{\cal L} =\sum_b {\cal L}_b + {\cal L}_m + {\cal L}_{\mathrm{int}} \ , 
\label{eq:lagrangian1}
\end{equation}
with
\begin{eqnarray*}
{\cal L}_b &=& \overline{\Psi}_b(i\gamma_{\mu}\overset{\xrightarrow{}}{\partial^{\mu}} - i\overset{\xleftarrow{}}{\partial^{\mu}} \gamma_{\mu} )\Psi_b - m_b\overline{\Psi}_b\Psi_b\ ,\\
{\cal L}_m &=& \frac{1}{2}\partial_{\mu}\sigma \partial^{\mu}\sigma - \frac{1}{2}m^2_{\sigma}\sigma^2 - b\sigma^3 - c\sigma^4 -\frac{1}{4}\Omega^{\mu \nu}\Omega_{\mu \nu}  \nonumber \\ 
&+& \frac{1}{2}m^2_{\omega}\omega_{\mu}{\omega}^{\mu} +\frac{1}{2}m^2_{\rho}\vec{\rho}_{\mu}\cdot\vec{\rho\,}^{\mu}-\frac{1}{4}P^{\mu \nu}P_{\mu \nu}\ ,\\
{\cal L_{\mathrm{int}}} &=& \sum_m \sum_b \frac{g_{mb}}{2}\left(\overline{\Psi}_b\overset{\xleftarrow{}}{D}_{mb}\Gamma_m\Psi_b \phi_m + \phi_m\overline{\Psi}_b\Gamma_m \overset{\to}{D}_{mb}\Psi_b\right) \ ,
\label{eq:lagrangian2}
\end{eqnarray*} 
where $\gamma_\mu$ are the Dirac matrices and the arrow indicates whether the action of the derivative operators is to be performed over $\overline{\Psi}_b$ (left) or $\Psi_b$ (right), 
with ${\Psi}_b$ being the field of baryon $b$ with mass $m_b$. The meson fields are represented by the symbols $\sigma, \omega$ and $\rho$, as well as by the generic notation $\phi_m$, corresponding to a meson $m$ with mass $m_m$. With $\Omega^{\mu \nu} =\partial^\mu\omega^\nu - \partial^\nu\omega^\mu$ and $P^{\mu \nu} = \partial^\mu\vec{\rho}^\nu - \partial^\nu\vec{\rho}^\mu$ we label the meson field tensors. In the interaction term, the Lorentz-factors $\Gamma_m = 1$, $\gamma^{\mu}$ and $\vec{\tau}\gamma^{\mu}$ act at the vertices involving scalar, vector and iso-vector mesons, respectively.\\ In the present work we adopt the constants of the SET I model presented in \cite{ChorozidouMomentumDependenceInMediumPotential2024}, which are tuned to the results of  the microscopic calculations employing $\chi$-EFT potentials \cite{PetschauerHyperonsChiralEFT2016}. As in Ref.~\cite{ChorozidouMomentumDependenceInMediumPotential2024}, we consider only $\Lambda$ and $\Sigma$ hyperons, since the uncertainty in the interactions of the $\Xi$ hyperons is still very large. In matter we consider electrons as the only leptons present and they are treated as free a Fermi gas.

The key feature of the NLD model is achieved by the additional regulator operators $\overset{\to}{D}$ that have a structure given by \cite{GaitanosAntinucleonOpticalPotential2011}
\begin{equation}
\overset{\to}{D} = \frac{\Lambda_2^2}{\Lambda_1^2\left[1+\displaystyle\sum_{j=1}^4 \left(\displaystyle \frac{v^{\mu}_j}{\Lambda_1}i\overset{\to}{\partial_{\mu}}\right)^2\right]},    
\end{equation}
where $v^\mu_j$ denote auxiliary vectors and $\Lambda_1$, $\Lambda_2$ are tunable cutoff parameters that depend on the species and the interaction channel. Within a mean field approximation, the complicated operator structure of $\overset{\to}{D}$ reduces to a momentum-dependent form factor:
\begin{equation}
D(p) = \frac{\Lambda_2^2}{\Lambda_1^2+\vec{p}^2} \ .    
\end{equation}
This form factor significantly changes the baryon properties in the medium. In particular the scalar and vector self energies
\begin{eqnarray}
    S_b(p) &=& g_{\sigma b} \sigma D_{\sigma b}(p) , \\
    V_b(p) &=& g_{\omega b}\omega D_{\omega b}(p)+\tau_{3b}g_{\rho b}\rho D_{\rho b}(p)  ,   
\end{eqnarray}
become now momentum dependent functions, which make the baryon effective mass:
\begin{equation}
    m^{*}_b(p) = m_b -S_b(p) , 
\end{equation}
and the baryon in-medium energy:
\begin{equation}   
E_{b}(p) = \sqrt{m^{*2}_b+p^2}+V_b(p) ,
\end{equation}
to acquire an additional momentum dependence that can alter the usual monotonic increase with momentum. 

We recall that hyperons are in weak equilibrium with the other species in neutron star matter, a fact that is imposed by the following relations between their chemical potentials and those of the proton, electron and neutron ($\mu_p, \mu_e$ and $\mu_n$): 
\begin{eqnarray}
&& \mu_{b^0} = \mu_n , \nonumber \\
&& \mu_{b^{-}} = \mu_n+\mu_e , \nonumber \\
&& \mu_{b^{+}} = \mu_p = \mu_n - \mu_e ,
\label{eq:chemical_potentials_relations}
\end{eqnarray}
for the neutral ($b^{0}$), negatively charged ($b^{-}$) and positively charged baryons ($b^{+}$), respectively. 

When the self-energy is momentum independent, the energy of the most energetic baryon, which at zero temperature is given by the chemical potential of the species, corresponds unequivocally to the particle having the highest momentum, which is denoted as the Fermi momentum $p_{F_b}$. In this situation, all momentum states below $p_{F_b}$ down to momentum zero are occupied with particles having energies smaller than the chemical potential of the species. The surface of this completely filled Fermi sphere is defined by the condition $E_b(p_{F_b})=\mu_b$. However, when the energy of the baryon, $E_b(p)$, is non-monotonic, the equation $E_b(p_{F_b})=\mu_b$ can have multiple solutions. As a result, the distribution of filled momentum states becomes more complex than a simple sphere, leading to an ``onion-like" structure where regions of filled and empty states alternate in momentum space. A proper account of this structure requires to reanalyze the condition for filled states within the NLD model. Indeed, the scalar and vector densities, $\rho_{b}^s$ and $\rho^v_{b}$, that act as source terms for the meson field equations, as well as the energy density and pressure of the species, involve integrals that are normally of the following form:
\begin{equation}
\label{eq1}
   \propto \int_0^{p_{F_b}}Q(p)\,dp \ ,
\end{equation}
where $Q(p)$ is the appropriate sub integral function that in the case of the scalar and vector densities includes the regulator $D(p)$. However, when the single particle energy of a baryon has a non-monotonic behavior with momentum, Eq.~(\ref{eq1}) does not hold anymore. Assume the equilibrium conditions of Eq.~(\ref{eq:chemical_potentials_relations}) establishes the chemical potential of baryon $b$ to be $\mu_b$, which corresponds to the maximum energy this baryon can have in matter at zero temperature, also referred to as Fermi energy. The non-monotonic single particle spectrum $E_b(p)$ can cross the constant line $\mu_b$ several times, and we label the momenta at the crossings as $p_{F_b}^{(i)}$, $i=1\dots n$. If the number of crossings $n$ is odd, then the integral in Eq. (\ref{eq1}) splits into:
\begin{equation}
    \int_0^{p_{F_b}}\longrightarrow \int_0^{p_{F_b}^{(1)}} + \int_{p_{F_b}^{(2)}}^{p_{F_b}^{(3)}}+ \cdots +\int_{p_{F_b}^{(n-1)}}^{p_{F_b}^{(n)}}.
\end{equation}
Similarly, if the number $n$ is even then the integral becomes:
\begin{equation}
    \int_0^{p_{F_b}}\longrightarrow  \int_{p_{F_b}^{(1)}}^{p_{F_b}^{(2)}}+\int_{p_{F_b}^{(3)}}^{p_{F_b}^{(4)}}+ \cdots +\int_{p_{F_b}^{{(n-1)}}}^{p_{F_b}^{(n)}}.
\end{equation}
We recall that at very large momentum ($p >\!> \Lambda_1,\Lambda_2$), the scalar and vector meson self energies tend to zero and $E_b(p)\propto p$. Therefore, the baryon spectrum will always overpass the threshold line $\mu_b$ at some large momentum, which ensures the even larger momentum states to be empty thus preventing the density of the baryon from diverging.

Having this fragmentation in mind, one can properly obtain the scalar and vector density of the baryons which, through the usual RMF equations, determine the meson fields (see the details of the procedure in Ref.~\cite{GaitanosMomentumDependentMFNuclearmatterNS2013}, for example). The particular composition of matter is obtained upon imposing the chemical equilibrium equations of Eq.~(\ref{eq:chemical_potentials_relations}), together with charge neutrality and baryon number conservation. 

Finally, the energy and the pressure can be obtained using the stress-energy tensor ~\cite{KochankovskiEoSHotHyperonicNSMatter2022}, taking into account the integral fragmentation .

As we will see in the next section, a proper treatment of the baryon occupation has a significant effect on the properties of dense matter. The possibility of having particles in matter at finite momentum, when they are absent at lower momenta, is a phenomenon already discussed in the context of matter created in heavy-ion collisions, where particles are maximally produced at non-zero momentum in the so-called ``moat" regime, i.e. when the single particle energy shows a minimum in the spectrum~\cite{PisarskiSignaturesMoatRegiveHIC2021,RenneckeParticleInterferometryMoatRegime}. We adopt the same nomenclature and will refer as ``moat hyperons" to those appearing at finite momentum in the absence of states at lower momenta.

\section{Results}
To clearly illustrate the effects arising from the treatment of the baryon spectrum in the medium, we perform three distinct types of beta-stable matter calculations at zero temperature. The first calculation, labeled ``$\mathrm{N}$", is a purely nucleonic one in which hyperons are artificially excluded from the composition of matter. The second calculation, denoted ``$\mathrm{NY}$  (no moat)", is performed in an analogous way to that in conventional RMF approaches, where the energy of a species increases monotonically with momentum. In this approach, a hyperon species appears at a certain onset density, when the condition $E_{\Lambda}(p=0) < \mu_n$ or $E_{\Sigma^-}(p=0) < \mu_n + \mu_e$ is fulfilled. The Fermi momentum of the hyperon species is determined by the solution of the equation $E(p_{F_Y}) = \mu_Y$ that is continuously connected to the solution at a slightly lower density, effectively ignoring the presence of other potential solutions (or "moats") that might arise from the non-monotonic energy-momentum dispersion relation. Finally, the ``$\mathrm{NY}$" calculation performs a consistent treatment of hyperonic matter, accounting for the potential appearance of moat particles.

\begin{figure}[ht]
\centering 
\includegraphics[width=0.48\textwidth, angle=0]{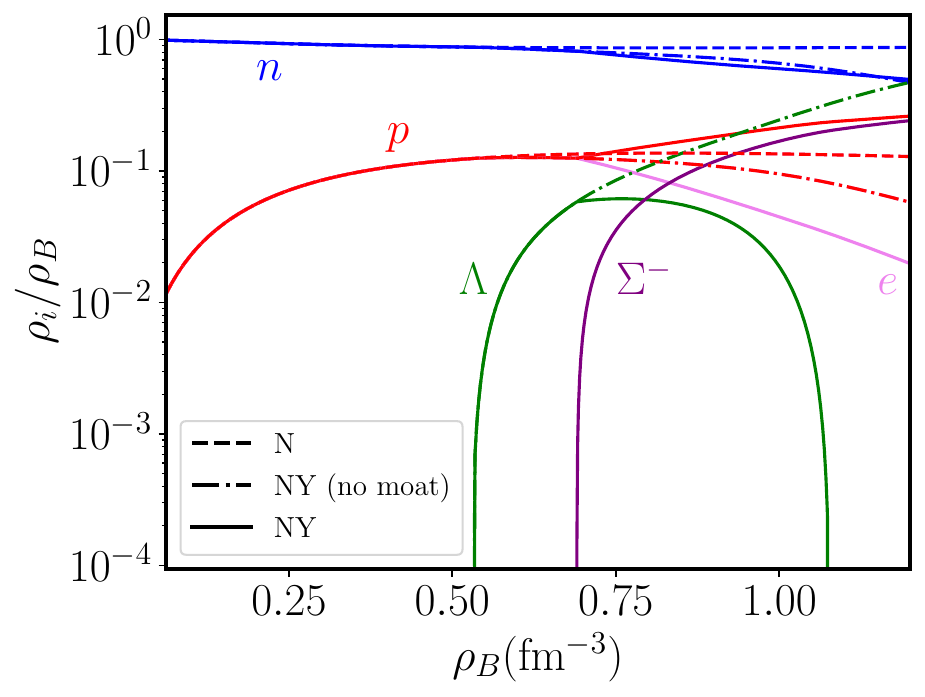}	
\caption{Composition of beta-stable matter. The different types of lines represent the three different calculations explored: taking only nucleons into account (dashed lines); considering hyperons but incorrectly treating the moat particles (dash-dotted lines); fully consistent NLD calculations (solid lines). Each color represents a different species. The abundance of the electron is different than that of the proton only when the $\Sigma^{-}$ hyperon is present in matter.} 
\label{Fig_composition}%
\end{figure}

Our results for the composition of beta-stable matter are presented in Fig.~\ref{Fig_composition}.
We find that when hyperons are excluded from matter (dashed lines), the particle abundances of neutrons, protons and electrons (the latter two being identical) slowly vary with increasing density. When hyperons are allowed to appear, their abundance is heavily influenced by the treatment of the Fermi sphere. Indeed, if one does not take into account the moat regime (dash-dotted lines), the $\Lambda$ is the only hyperon species present in matter. Its abundance continuously increases with density, and it can even surpass that of the neutron at around $7-8$ times saturation density. However, if the fragmentation of the Fermi sea is treated consistently and the presence of moat hyperons is considered (solid lines), the composition changes significantly. Indeed, we can see that right below $\rho_B\approx0.75~\mathrm{fm^{-3}}$, the $\Sigma^{-}$ hyperon starts to be energetically allowed. 
Its appearance immediately reduces the abundance of energetic electrons, consequently decreasing the neutron population and hence the amount of $\Lambda$ hyperons, which quickly vanish from matter. As the abundance of $\Sigma^-$ hyperons increases with density, so does that of protons due to charge neutrality.

\begin{figure}[ht]
	\centering 
	\includegraphics[width=0.50\textwidth, angle=0]{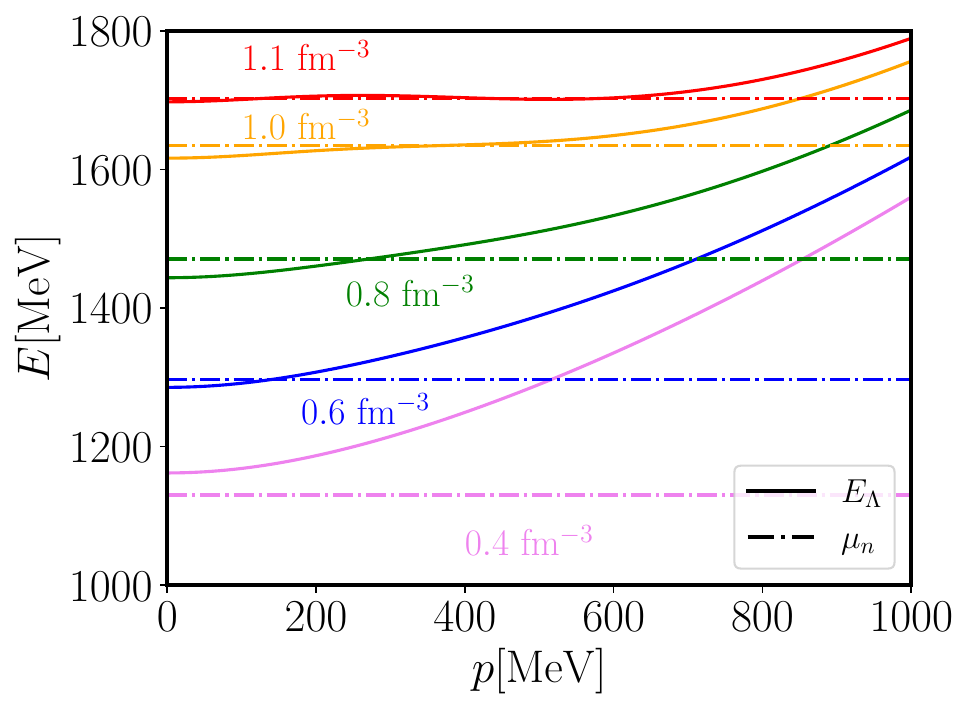}	
	\caption{ $\Lambda$ spectrum in beta-stable matter as a function of the momentum, for various densities within the ``$\mathrm{NY}$~(no moat)" approach. The horizontal dash-dotted lines represent the chemical potentials $\mu_n$.} 
	\label{fig:lambda_no_moat}%
\end{figure}
To better understand the vastly different composition patterns it is convenient to inspect the behavior of the hyperon spectra in matter.
We first focus our attention on the composition obtained within the ``$\mathrm{NY}$  (no moat)" approximation and we compare it with the results of Ref.~\cite{ChorozidouMomentumDependenceInMediumPotential2024}. We find our baryon composition to be consistent with the results displayed in Fig.~3(a) of Ref.~\cite{ChorozidouMomentumDependenceInMediumPotential2024}\footnote{We are employing the parameters that give rise the upper limit of the EoS band shown in Fig.~3(c) of Ref.~\cite{ChorozidouMomentumDependenceInMediumPotential2024}.} up to $\rho_B \sim 0.95$~fm$^{-3}$. The $\Lambda$ hyperons abruptly disappear from matter at this density in that work, while their amount keeps increasing with density in our case. The abrupt disappearance of hyperons is difficult to understand, especially when the $\Lambda$ hyperon momentum-dependent spectrum lies below the neutron chemical potential for a broad range of momenta, as seen in the $\rho_B=1.0$~fm$^{-3}$ panel of Fig.~2 in~\cite{ChorozidouMomentumDependenceInMediumPotential2024}. The results shown there suggest the existence of a Fermi sea of $\Lambda$ hyperons with a very large Fermi momentum, of about 3.5~fm$^{-1}$, leading to a $\Lambda$ density which is higher than the baryonic density quoted for that panel. 

To help in analyzing the inconsistencies that arise when the moat region is ignored,  we show in Fig.~\ref{fig:lambda_no_moat} the energy of the $\Lambda$ hyperon as a function of momentum (solid lines) together with the neutron chemical potential $\mu_n$ (dashed-dotted lines), for different densities (different colours), obtained within the ``$\mathrm{NY}$  (no moat)" approach \footnote{In this approach, which disregards the moat region, the $\Sigma^-$ hyperons do not appear and they are not relevant for the present discussion.}. At low densities, the $\Lambda$ spectrum lies above $\mu_n$ so neutrons cannot convert to $\Lambda$ hyperons, which are then absent in matter. As density increases so does $\mu_n$ and eventually it crosses the $\Lambda$ spectrum once, at the position of a well-defined $\Lambda$ Fermi momentum, $p_{F_\Lambda}$. The situation becomes especially interesting at the highest density displayed, where $\mu_n$ crosses the $\Lambda$ spectrum three times, leading to two regions of occupation, from 0 to $p_{F_\Lambda}^{(1)}$, and from $p_{F_\Lambda}^{(2)}$ to $p_{F_\Lambda}^{(3)}$ (the latter is the moat region). In the ``$\mathrm{NY}$  (no moat)" approach we disregard the moat region and assume all momentum states occupied from 0 to $p_{F_\Lambda}^{(2)}$. By not  considering the fragmentation of the $\Lambda$ Fermi sphere  properly in the present ``$\mathrm{NY}$  (no moat)" approach,  we obtain a population of $\Lambda$ hyperons that keeps increasing with density, but we cannot explain their abrupt disappearance as claimed in Ref.~\cite{ChorozidouMomentumDependenceInMediumPotential2024}.

\begin{figure*}[ht]
	\centering 
	\includegraphics[width=0.99\textwidth, angle=0]{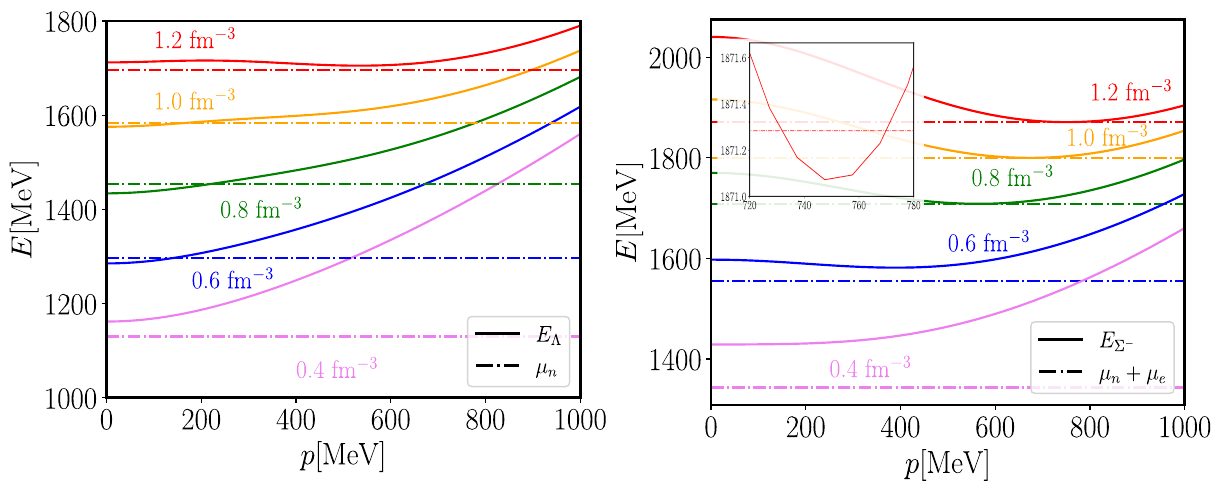}	
	\caption{Spectra of the $\Lambda$ (left panel) and $\Sigma^{-}$ hyperon (right panel) in beta-stable matter as a function of the momentum of the particles for various densities within our consistent "NY" calculations. The horizontal dash-dotted lines represent the chemical potentials $\mu_n$ (left plot) and $\mu_n+\mu_e$  (right plot). 
    On the right plot, the zoomed version of the $\Sigma^{-}$ moat region, for density $\rho_B = 1.2~\mathrm{fm^{-3}}$, is also shown.} 
	\label{InMediumEnergiesLambdas}%
\end{figure*}

When the particle occupation is properly taken into account in our consistent ``NY" calculation, we obtain the hyperon spectra (solid lines) shown in 
Fig.~\ref{InMediumEnergiesLambdas}, together with the conditions for their appearance, i.e. $\mu_n$ for the $\Lambda$ hyperons on the left panel and $\mu_n+\mu_e$ for the $\Sigma^-$ hyperons on the right panel.  

Similarly as for the case of the ``$\mathrm{NY}$  (no moat)" approach shown in Fig.~\ref{fig:lambda_no_moat}, the $\Lambda$ hyperons are absent at lower densities until they eventually appear, and subsequently increase their population with density. However, in the present consistent calculation, the presence of the moat $\Sigma^-$ hyperons at a certain density induces a decrease of $\mu_n$ and the crossing with the $\Lambda$ spectrum happens at a lower momentum,  hence reducing the $\Lambda$ population from that density onwards. Eventually, at density $\rho_B=1.2~\mathrm{fm}^{-3}$ there is no longer a crossing, explaining why in the composition plot of Fig.~\ref{Fig_composition} the $\Lambda$ hyperons have disappeared at these high densities. 

The situation for the $\Sigma^{-}$ hyperon is completely different. The lack of crossing between the $\Sigma^{-}$ spectrum (solid lines) and the threshold energy $\mu_n + \mu_e$ (dash-dotted lines) at the lower densities, explains the absence of $\Sigma^{-}$ hyperons. However, at a density of $\rho_B = 0.8~\mathrm{fm^{-3}}$ the two lines do cross and the $\Sigma^-$ hyperon is present in matter. Contrary to the case of the $\Lambda$ hyperon at the same density, due to the non-monotonic behavior of the $\Sigma^-$ spectrum in the medium, the minimum is not found at $p=0$, but at a fairly large momentum of $p \gtrsim  500~\mathrm{MeV}$. Note that all $\Sigma^{-}$ hyperons present in matter at this density have now finite momenta, with values within the range established by the two crossings between the solid and the dash-dotted lines. These are the moat hyperons. The low momentum states are too energetic to be populated. The same pattern continues at higher densities, only shifting the relatively shallow moat region to higher momenta. The inset on the right panel of Fig.~\ref{InMediumEnergiesLambdas} shows a zoomed version of the moat region of $\Sigma^{-}$ hyperons for a density $\rho_B = 1.2~\mathrm{fm^{-3}}$. One can see there that the energies of the least and most energetic $\Sigma^{-}$ hyperon differ only by about $0.2~\mathrm{MeV}$! This suggests that the energy of the $\Sigma^{-}$ hyperons in beta stable matter at a fixed density is almost constant, while their momenta cover a thin layer of the Fermi sphere of less than 50 MeV width. Even in this situation, the amount of $\Sigma^-$ hyperons can be very large at high density, reaching an abundance of 20\% at $\rho=1.2~\mathrm{fm^{-3}}$.

\begin{figure}[ht]
	\centering 
	\includegraphics[width=0.48\textwidth, angle=0]{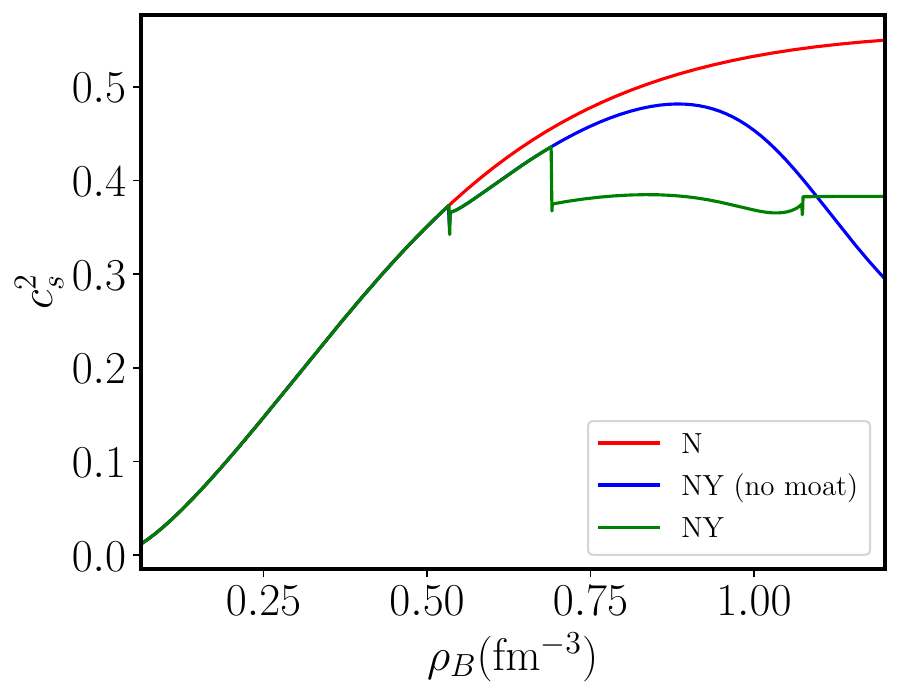}	
	\caption{Speed of sound in beta-stable matter. Different line colors represent the three different calculations: red line - calculations taking into account only nucleons; blue line - calculation without taking into account the moat particles; green line - full consistent NLD calculation.} 
	\label{SpeedOfSound}%
\end{figure}

The differences in the composition patterns from the different approximations is even better seen if one inspects, in Fig.~\ref{SpeedOfSound}, our results for the speed of sound, $c_s$, which is obtained from the derivative of the total pressure in matter $P$ with respect to the total energy density $\varepsilon$ 
\begin{equation}
    c_s^2 = \frac{dP}{d\varepsilon}.
\end{equation}
In the absence of hyperons, the speed of sound is a monotonically increasing function of $\rho_B$. This is no longer the case when hyperons are present in beta-stable matter. At the density where $\Lambda$ hyperons start appearing, the speed of sound shows a drop. This is an expected result, and it is a consequence of the softening of the EoS that hyperons produce in matter \cite{MottaSoundSpeedHyperonicStars2021,AguirreHyperonsSpeedOfSound2022}. However, the speed of sound immediately recovers for both cases (``$\mathrm{NY}$  (no moat)" and ``$\mathrm{NY}$") being only marginally lower than that obtained in the absence of hyperons. This is a unique feature of the NLD model, as we explain in what follows. We recall that, even when hyperons are not considered in the EoS, the nucleonic contributions can make the EoS softer at higher densities due to the form-factor regulator that inhibits the sources of the meson fields at high momenta. When $\Lambda$ hyperons appear at zero momentum, they replace the more energetic neutrons and, while the $\Lambda$ hyperons soften the EoS, they are also responsible for removing high momentum neutrons that, due to the regulator, can have a relatively soft behavior. The tread-off between these two effects yields a hyperonic EoS with a speed of sound that closely resembles the nucleonic one within a particular density region. However, this trend is broken in the two hyperonic models at higher densities. When moat particles are not considered,  the speed of sound starts to decrease at sufficiently high densities, where the abundance of $\Lambda$ hyperons is significant. This is a consequence of the lower cutoff (stronger momentum softening) of the $\Lambda$ hyperons compared to that of the nucleons. When the presence of moat hyperons is consistently taken into account, the speed of sound experiences a significant drop at the density where the $\Sigma^{-}$ hyperons appear. From that drop onwards, the speed of sound stays flat. This is because these hyperons are produced at finite (and rather high) momenta that suffer an additional softening from the regulator, hence preventing the speed of sound from increasing significantly. The only noticeable feature is a small spike in the speed of sound at very large densities, signaling the density at which the $\Lambda$ hyperons disappear from matter.

\begin{figure}[ht]
	\centering 
	\includegraphics[width=0.48\textwidth, angle=0]{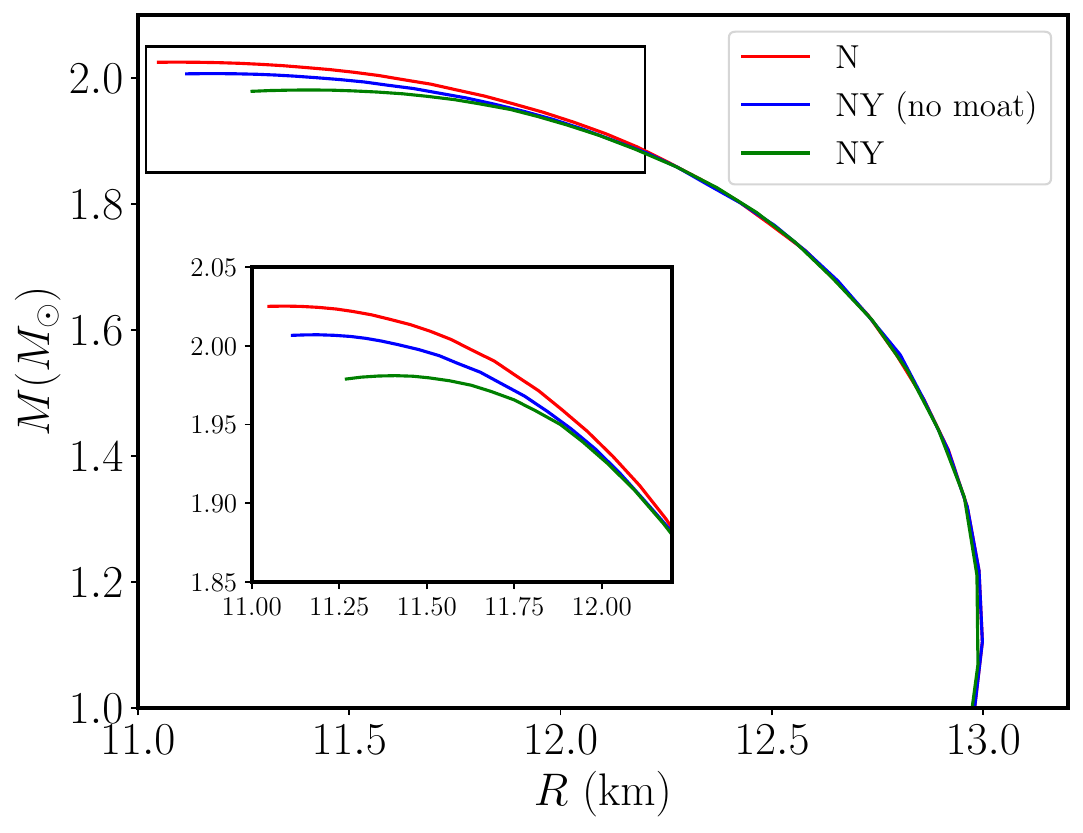}	
	\caption{$M(R)$ in beta-stable matter calculated with our three different EoSs (red line - calculations taking into account only the nucleons; blue line - calculation without taking into account the moat particles; green line - full consistent NLD calculation). We also show the region of most massive stars in a zoomed form embedded in the plot.} 
	\label{MR}%
\end{figure}

Although these effects seem significant at the EoS level, they do not translate into widely different mass-radius curves, $M(R)$\footnote{We used the software \cite{DavisSoftwareCUTER2024}, presented in \cite{DavisInferenceNSPropertiesUnifiedCrustCoreEoS2024}, to obtain a crust consistent with our high-density EoS. }, \cite{OppenheimerMassiveNCTOV1939} as seen in Fig.~\ref{MR}. 
We observe that the $\Lambda$ hyperons are present in the core of stars with masses larger than $M\gtrsim 1.5~M_{\odot}$. However, the $M(R)$ relations of the hyperonic models do not deviate from the nucleonic one until reaching a mass of $M \sim 1.9~M_{\odot}$, which corresponds to a central density inside hyperonic stars of around $\rho_{B,c}\approx 0.7~\mathrm{fm^{-3}}$. For those stars, the $\Sigma^{-}$ hyperons are not yet present in the core. When they eventually appear in the consistent $``\mathrm{NY}"$ calculation (green line) for stars with masses $M\gtrsim 1.93M_{\odot}$, the $M(R)$ relation  quickly reaches its maximum mass of around $M\sim 1.98M_{\odot}$. The nucleonic only model (red line), on the other hand, predicts a maximum mass of around $M\sim 2.03M_{\odot}$. We can conclude that the $M(R)$ observations of low- and intermediate-mass stars could not help to learn about the composition of matter within the NLD model.

\section{Summary and conclusions}
In this work, we have revisited the hyperonic EoS of dense matter employing the non-linear derivative model within the relativistic mean-field approach. The main feature of this model is the consideration of a derivative operator, $\overset{\to}{D}$, which, in homogeneous nuclear or hypernuclear matter, simplifies to a monopole-type regulator function that softens the interaction at high momenta. This softening effect can alter the otherwise monotonic behavior of the energy of the baryonic species as a function of momentum.

We explicitly demonstrated that the hyperonic species present in matter, specifically $\Lambda$ and $\Sigma^{-}$ hyperons, can have a non-monotonous single particle spectrum. In ultra-dense matter, the minimum energy can be achieved at a finite momentum, allowing states to be populated even when regions of lower momenta are empty. In such cases, the usual Fermi sphere is disrupted, acquiring an onion-like distribution which alternates empty and filled momentum regions.
This is a similar phenomenon to the ``moat region" predicted to appear in heavy-ion collisions. Drawing an analogy with this terminology, we refer to these interstitial particles as moat hyperons.

The appearance of moat hyperons significantly influences the composition patterns and the EoS of the matter. Due to the softening of the interaction at higher momenta induced by the regulator, the presence of moat hyperons can prevent a significant increase in the speed of sound, which is also an indication of their softening effect on the EoS. However, as these particles emerge only in ultra-dense matter, they do not substantially modify the $M(R)$ relation for low and intermediate mass neutron stars.

 Hence, the main message of this work is to underscore the importance of considering the momentum dependence of the interaction in dense matter, particularly when exotic species such as hyperons are present. Furthermore, we emphasize the necessity for a proper treatment of the fragmentation of the Fermi sphere when dealing with momentum-dependent interactions in hypernuclear matter. The methodology we have outlined here can be applied regardless of the specific framework used to obtain the hyperonic EoS.

Finally, we want to highlight the ongoing need for clear observational signatures from neutron stars that contain hyperons. These signatures could help resolving the long-standing question about the presence of hyperons in neutron star matter and significantly constrain the baryon in-medium interactions. As demonstrated in this work, the cold star observables can be rather degenerate with respect to the composition, raising the importance of exploring other observational avenues. In recent years, several works have been done in this direction, both at zero and finite temperature \cite{BanikProbingMetastabiliyPNS2014, MalikBayesianInferenceHyperons2022, GModeOscillationsNSHyperons2023, BlackerThermalBehaviorHyperonsBNSM2024, KochankovskiImpactHyperonsNSM2025, ZhengPreprintFModeOscillationsPNS2025, FischerHyperonsPNSDeleptonizationDarkFlavouredParticles2025,BarmanFModeOscillationsHotNSHyperons2025,FerreiraIdentifyingHyperonsNSMRSlope2025}. Notably, observations of binary neutron star mergers could be sensitive to the thermal behavior of hyperonic matter, which has been predicted to be rather different from the one of nucleonic-only matter \cite{KochankovskiEoSHotHyperonicNSMatter2022,RadutaEoSHotNeutronStarsExoticParticle2022}. Therefore, an extension of this model to finite temperatures would be highly beneficial to investigate whether the finite-temperature hyperonic characteristics are also reproduced when a more complex and momentum-dependent model is employed.

\section*{Acknowledgments} 
We thank Jürgen Schaffner-Bielich, Arnau Rios, Mario Centelles and Xavier Roca Maza for the useful discussions. This research has been supported from the projects CEX2024-001451-M, CEX2020-001058-M (Unidades de Excelencia ``Mar\'{\i}a de Maeztu"), PID2023-147112NB-C21 and PID2022-139427NB-I00 financed by MCIN/AEI/10.13039/501100011033/FEDER, UE, as well as by the EU STRONG-2020 project under the program H2020-INFRAIA-2018-1 grant agreement no. 824093. H.K. acknowledges support from the PRE2020-093558 Doctoral Grant of the spanish MCIN/ AEI/10.13039/501100011033/. L.T. also acknowledges support from the Generalitat Valenciana under contract CIPROM/2023/59, from the Generalitat de Catalunya under contract 2021 SGR 171, and from the CRC-TR 211 'Strong-interaction matter under extreme conditions'- project Nr. 315477589 - TRR 211. 

\bibliographystyle{elsarticle-num} 
\bibliography{example}

\end{document}